# Applications of AI in Astronomy

S. G. Djorgovski*, A. A. Mahabal*, M. J. Graham*, K. Polsterer[†],
A. Krone-Martins[‡]

**Abstract:** We provide a brief, and inevitably incomplete overview of the use of Machine Learning (ML) and other AI methods in astronomy, astrophysics, and cosmology. Astronomy entered the big data era with the first digital sky surveys in the early 1990s and the resulting Terascale data sets, which required automating of many data processing and analysis tasks, for example the star-galaxy separation, with billions of feature vectors in hundreds of dimensions. The exponential data growth continued, with the rise of synoptic sky surveys and the Time Domain Astronomy, with the resulting Petascale data streams and the need for a real-time processing, classification, and decision making. A broad variety of classification and clustering methods have been applied for these tasks, and this remains a very active area of research. Over the past decade we have seen an exponential growth of the astronomical literature involving a variety of ML/AI applications of an ever increasing complexity and sophistication. ML and AI are now a standard part of the astronomical toolkit. As the data complexity continues to increase, we anticipate further advances leading towards a collaborative human-AI discovery.

## 1.1 Introduction and Background

Astronomy entered the era of big data with the advent of large digital sky surveys in the 1990s, which opened the TB-scale regime. Sky surveys have been the dominant source of data in astronomy ever since, reaching the multi-PB scale by the late 2010s; see, e.g., [1] for a review. This stimulated the creation of the Virtual Observatory (VO) framework [2,3], which has now evolved into a global data grid of astronomy (see https://ivoa.net), providing access to data archives from both ground-based observatories and surveys, and the space-based missions.

This wealth and growth of data rates, volumes, and complexity demanded applications of automated data processing and analysis tools. While VO and the individual archives provide data access and some tools, most of the remaining astronomical cyberinfrastructure and data analytics tools have been developed by the individual research groups, captured under the Astroinformatics umbrella. Today, applications of Machine Learning (ML) and other AI methods are becoming commonplace and growing rapidly. During the 2021, according to the Astrophysics Data System (ADS; https://ui.adsabs.harvard.edu), there were about 1000 astronomy/astrophysics papers that involved ML or AI, and their numbers are growing exponentially, with a doubling time of about 20 months. AI is now a standard part of the astronomical toolkit.

---

* California Institute of Technology, Pasadena, CA 91125, USA

[†] Heidelberg Institute for Theoretical Studies, 69118 Heidelberg, Germany

[‡] University of California, Irvine, CA 92697, USA

ML/AI methods are used to create value-added higher level data products for follow-up research, and may include source detection and segmentation tasks, structural and morphological classification, as well as all kinds of ordinary classification and regression tasks. While supervised classification tools are by construction unable to detect any new types of objects that are not present in the training data sets, unsupervised clustering offers a possibility of discovering previously unknown classes, and enable detection of rare, unusual, or even previously unknown types of objects as outliers in some feature space.

Given the vast scope of the field, the goal of this chapter is not to provide a comprehensive review, but rather to give some illustrative examples where ML/AI have enabled significant advances in astronomy.

A useful didactic overview is by [4]. The monograph [5] gives an extensive and practical coverage of the subject. For some recent examples, see [6].

## 1.2 Early Applications: Digital Sky Surveys

While there have been some early experiments in the early 1990s, the use of ML/AI in astronomy really started growing in mid-1990s, with the advent of the first large digital sky surveys [7,8]. The initial applications were to automate repetitive tasks that were previously done by humans. A good example of the star-galaxy separation for the catalogs of objects detected in sky surveys. Following the image segmentation that identifies individual sources, several tens to hundreds morphological and structural parameters are evaluated for each one, thus forming feature vectors that can be analyzed using ML tools. Essentially, first Terabytes and now Petabytes of images are converted into database catalogs of many millions to billions of feature vectors, each representing an individual detected source, in data spaces of hundreds of dimensions, or even thousands once multiple catalogs are combined.

In the visible, UV, and IR wavelength regimes, the first order classification is between the unresolved sources ("stars") and resolved ones ("galaxies"), based purely on the image morphology. Supervised classification methods, such as the Artificial Neural Networks (ANN), or Decision Trees (DT) were used effectively for this task; see, e.g., [9,10]. Morphological classification can be used to identify and remove a variety of instrumental artifacts, that may appear as outliers in the feature space, [11]. The physical classification as objects of different types, e.g., Galactic stars vs. quasars, different types of galaxies, different types of stars, etc., requires additional data in a subsequent analysis.

Digital sky surveys opened a new, highly effective mode of astronomical discovery. Traditional mode of pointed observations focuses on the individual objects or small samples of objects. Sky surveys may detect billions of sources, and ML can be used to select and prioritize the most interesting targets for the follow-up with large and/or space-based telescopes, thus optimizing the use of these scarce resources. Applying supervised classifiers or cuts in the feature space informed by the domain expertise has been proved to be very effective in finding well-defined samples of relatively rare types of objects, such as high-redshift quasars or brown dwarfs [12,13,14].

**1.3 Increasing Challenges: Time-Domain Astronomy**

Improvements in the size, quality, and cost of astronomical detectors enabled much larger format imaging cameras, which in turn enabled the rise of the synoptic sky surveys, where large areas of sky are surveyed repeatedly. This opening of the Time Domain enabled systematic, scaled-up studies of various temporal phenomena, including variability of stars and active galactic nuclei, cosmic explosions of all kinds (e.g., many types of Supernovae, Gravitational Wave events, etc.), moving objects such as the potentially hazardous asteroids, etc. Time Domain – essentially a panoramic cosmic cinematography – touches all fields of astronomy, from the Solar System to cosmology.

This opens new scientific opportunities, but also brings new challenges in addition to those posed by the traditional sky surveys, by adding the variability information to the classification tasks, and often time-criticality due to the transitive nature of the observed events: the potentially most interesting ones have to be identified in real time, as they must be followed up before they fade away [14,15,16,17,18,19,63,64]. Figure 1 shows a conceptual illustration of some of these challenges.

In addition to the full electromagnetic (EM) spectrum, gravitational wave, high energy cosmic rays, and neutrino observatories are also now providing copious amounts of data, opening the field of Multi-messenger Astronomy. In general, the events detected by these non-EM channels have a very poor angular resolution and directionality but identifying their EM counterparts is essential for their physical interpretation. This leads to large area searches for counterparts, with many potential candidates. ML methods can be used to classify the possible candidate counterparts, e.g., [20,63,65,66], and other ML methods are being used to scrutinize and classify the non-EM signals themselves, e.g., [21,22].

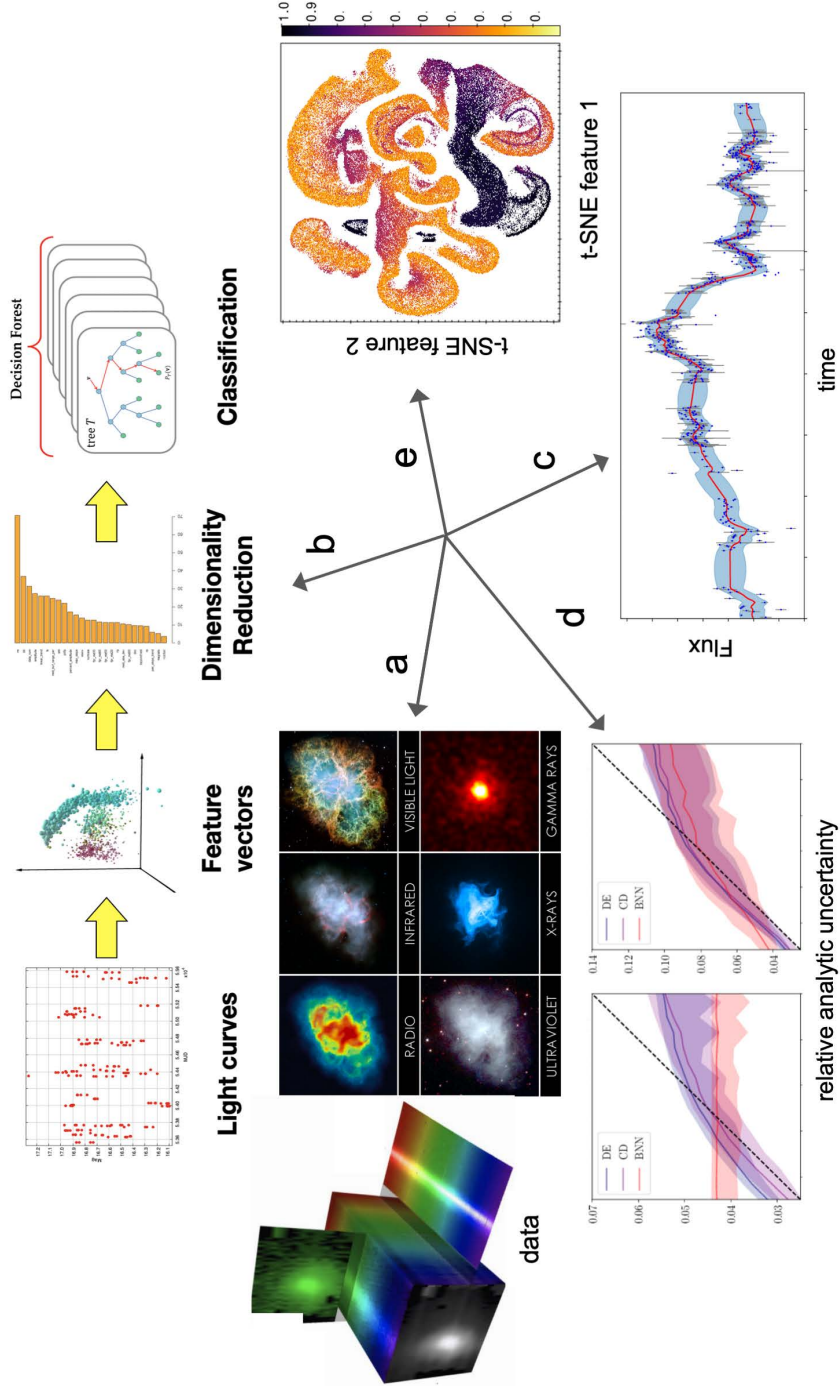

**Figure 1**: The five major challenges facing TDA today (a) a representation of a data cube signifying multiple wavelengths and multiwavelength observation of the Crab Nebula, (b) multiplicity of data features, (c) GPRs as a way to bridgegaps in observing, (d) Understanding effect of noise on objects (from Caldeira & Nord 2020), and (e) detecting anomalies (from Martinez-Galarza et al. 2021). Image credits: João Steiner, Stephen Todd, ROE and Douglas Pierce-Price, JAC for datacube; NRAO/AUI and M. Bietenholz; NRAO/AUI and J.M. Uson, T.J. Cornwell (radio); NASA/JPL-Caltech/R. Gehrz / University of Minnesota (infrared); NASA, ESA, J. Hester and A. Loll / Arizona State University (visible); NASA/Swift/E. Hoversten, PSU (ultraviolet); NASA/CXC/SAO/F. Seward et al.(X-rays); NASA/DOE/Fermi LAT/R. Buehler (gamma rays) for Crab Nebula images.

Predicting the value of a time series between measurements is a common problem, particularly with multidimensional time series where values for all quantities at all timestamps are not available. In astronomy, this can pertain to prioritizing follow-up observations for interesting transient events. In the absence of a good theoretical model for the underlying process, a variety of interpolation schemes can be employed. However, these can also be a source of additional systematic errors.

One critical issue in classification in general is the poor scaling of the classification algorithms with the dimensionality of the feature space. This is especially important in the Time Domain; some approaches have been discussed by [23,24]. Feature dimensionality from several tens to few hundreds are now routine. Not all of these are independent (orthogonal) and some even add to noise for specific classes, making dimensionality reduction a critical need. For multiclass classification, disambiguating features that are important for one class but not another can be non-trivial. An example is the use of binary classifiers for the Zwicky Transient Facility (ZTF) [25]. Tools like DNN and XGBoost can be also used to identify the top features. When external features (e.g., from crossmatches to other surveys) are being incorporated, there can be missing features, which further complicates the use of some of the ML techniques. Techniques like Data Sheets [26] and Model Cards [27] can be used to standardize data set creation and lead to reusable models [28].

An alternative is to adopt a probabilistic approach and regard the observed time series as a single draw from a multivariate Gaussian distribution fully characterized by a mean (normally assumed to be zero or constant) and a covariance function. Predicted points then just represent a subsequent draw and can be calculated with an associated uncertainty. Gaussian process regression (GPR) [31] uses the observed data to learn the hyperparameters of the underlying covariance kernel (and mean function if present). The form of the kernel function is a choice for the user but specific properties of the underlying process, such as stationarity, autoregressive behavior, or a spectral density representation, are easily represented. In fact, there is a theoretical duality between certain GPR kernels and neural network architectures [32] and neural networks can also be employed directly as kernels in some GPR implementations.

The main issue in using GPR, particularly with large multidimensional data sets with thousands of data points and/or comprising thousands of samples, is the speed of fitting since this typically involves extensive matrix inversion. Certain GPR implementations have achieved good performance by making very specific decisions on the functional form of the kernel function, e.g., it only uses exponential terms, and optimizing accordingly, but this is not a global solution [33].

It is also unclear how good GPR is at forecasting data. There is a tendency with non-periodic kernels to regress to the mean with increasing time past the last observed data point and increasing uncertainty. Recurrent neural networks seem to show better performance [34], but they have also been used more in forecasting a next data point rather data over an extended period of time. Obviously, if the underlying process is nonlinear, then forecasting presents a much bigger problem and more advanced deep learning architectures may be required.

Incorporating error-bars in ML analysis is a notoriously non-trivial challenge. That fact, combined with classifying objects near the detection limit, raises a very different kind of a challenge. Observing ever-fainter objects helps push the science into newer areas, but at the same time it comes with a greater risk of such objects turning out to be false positives (e.g., short lived transients that are below the quiescent limit that brighten into observable range for a short duration). Uncertainty quantification, e.g., [35], combined with emulators and simulators [29], along with models based on Bayesian techniques may be needed [36].

Looking for anomalies, i.e., objects or events that do not seem to belong to any of the known classes, poses additional challenges. It is difficult to define what an anomaly is in the first place: is it an out-of-distribution object, a completely independent or new class? Will more observations favor one interpretation or the other, and what kind of observations will be required? Active learning where newer observations are iteratively used to improve classification routinely indicate the requirement to revise classifications at the boundaries of ambiguity. However, the follow-up resources are limited, and with the ever-growing data rates it becomes critical to optimize their use. Some early efforts in this arena include [30,37,38,39,40], and others.

### 1.4 A Growing Applications Landscape

While classification and outlier search remail a staple of survey-based astronomy, there is now a much broader range of ML/AI applications.

One application area of ML tools is the estimation of photometric redshifts (photo-z's). In cosmology, redshift, reflecting the increase in the scale of the universe since the light was emitted by some distant object, due to the cosmic expansion is a measure of distance, and thus necessary in determining the physical parameters of distant galaxies and quasars, such as their luminosities, masses, etc. Traditionally, redshifts are measured spectroscopically, which is observationally expensive. Photo-z's are a way of estimating the redshifts from the multi-color photometry, which is observationally much cheaper, and thus can be done for vastly larger numbers of objects. This is a use case where ML has enabled significant advances and savings of observing time, by including instrumental characteristics and physical models implicitly expressed through the data. There is an extensive literature on this subject with a wide variety of ML methods [41,42,43,44,45,46,47,48,49,50], and many others). One important conceptual change was to replace a single number representation of an estimated photo-z with a probability density distribution.

ML has been also used as a method for the discovery of gravitational lenses, operating both on the images from sky surveys [51], and time series of quasar variability [52]. Such large, systematically selected samples of gravitational lenses can be used as probes of dark matter and the expansion rate of the universe.

In addition to the data obtained through observations, the output of large numerical simulations is another large source of data calling for an automated analysis. Current developments led to hybrid approaches of classically computing physical models in combination with ML models. By doing so, time-consuming calculations can be replaced through very quick ML models that act as a surrogate with a similar accuracy as the original code. Thereby, larger volumes, with higher spatial and temporal resolution, can be computed without requiring more resources. Another aspect is that critical branching conditions can be detected, and the resolutions of the simulation can be adaptively changed, ensuring that those details are not lost or overseen while simulating. Some examples include [53,54,55,56], and others.

This is just a small sampling of the examples of the diverse, growing, and creative uses of AI/ML in astronomy.

### 1.5 Concluding Comments and Future Prospects

While the range and diversity of AI applications in astronomy continue to grow, most applications so far have been focused on the analysis of already acquired data sets and their derived data products, such as the tables of pre-extracted and expert engineered features. However, the landscape of ML/AI applications is also changing rapidly, affecting different stages of a data-centric approach, data acquisition, processing, and analysis.

The acquisition stage covers the process of planning and performing observations. For the most part this process was done by the individual experts, but based on specified quality criteria, AI systems can learn to perform this planning automatically. Based on the quick analysis of the initial data, instrument setup, exposure times, etc., can be used to close the loop between choosing the right observational setup and getting high quality science data. This kind of fast, AI-based observation planning and control systems will likely replace the current way of planning, scheduling, and observing, leading to an improved scientific outcome, both in quality and quantity of the observations. This will be critical in the arena of Time-Domain and Multi-Messenger Astronomy, where transient events may be detected and followed up by a number of different surveys and facilities, and their prioritization for the follow-up would be essential.

Incorporation of domain knowledge into ML/AI analysis, such as the 'physics-based' AI is an active area of research, with many outstanding challenges remaining. Some examples include [57,58], and others.

Besides the analysis of scientific data, ML methods get utilized to access complex scientific content like scientific publications or to realize a natural language and chat-based access to data stored in catalogs. ML and AI based systems may transform the way of finding and accessing data soon. Likewise, ML can be used to sort through the literature given a set of user preferences; an example is http://www.arxiv-sanity.com/.

Another novel direction is in using AI not just to map the data spaces and find interesting objects, but to discover potentially interesting relationships that may be present in the data. One approach is to use symbolic regression, e.g., [59,60]. A related technique uses memetic regression [61,62].

As the data complexity continues to increase, use of AI to detect interesting patterns or behaviors present in the data, that may elude humans, e.g., due to the hyper-dimensionality of the data will keep on increasing. The interpretation of such AI-based discoveries still rests with the humans, but there is a possibility that some of them may simply exceed the human cognitive capabilities. We may increasingly see more examples of a collaborative human-AI discovery.

# References

[1] Djorgovski, S. G., Mahabal, A. A., Drake, A., Graham, M. J., and Donalek, C. (2012). Sky Surveys, in: Astronomical Techniques, Software, and Data, ed. H. Bond, Planets, Stars, and Stellar Systems, **2**, ser. ed. T. Oswalt, (Dordrecht: Springer), pp. 223-281.

[2] Brunner, R. J., Djorgovski, S. G., and Szalay, A. S. (2001). Virtual Observatories of the Future, A.S.P. Conf. Ser. **225** (San Francisco: Astronomical Society of the Pacific).

[3] Djorgovski, S. G., and the NVO Science Definition Team (2002). Towards the National Virtual Observatory, https://www.nsf.gov/mps/ast/sdt_final.pdf (Washington, DC: National Science Foundation).

[4] Baron, D. (2019). Machine Learning in Astronomy: A Practical Overview, available online at https://arxiv.org/abs/1904.07248

[5] Ivezić, Ž., Connolly, A., VanderPlas, J., and Gray, A. (2020). Statistics, Data Mining, and Machine Learning in Astronomy: A Practical Python Guide for the Analysis of Survey Data. (Princeton: Princeton University Press).

[6] Zelinka, I., Brescia, M., and Baron, D. (editors) (2021). Intelligent Astrophysics, In: Emergence, Complexity and Computation, **39,** Springer Nature.

[7] Weir, N., Fayyad, U., Djorgovski, S. G., and Roden, J. (1995a). The SKICAT System for Processing and Analysing Digital Imaging Sky Surveys, Publ. Astron. Soc. Pacific **107**, 1243-1254.

[8] Fayyad, U., Smyth, P., Weir, N., and Djorgovski, S. G. (1995). Automated Analysis and Exploration of Image Databases: Results, Progress, and Challenges, J. Intel. Inform. Sys. **4**, 7.

[9] Weir, N., Fayyad, U., and Djorgovski, S. G. (1995b). Automated Star/Galaxy Classification for Digitized POSS-II, Astron. J. **109**, 2401-2414.

[10] Odewahn, S., de Carvalho, R., Gal, R., Djorgovski, S. G., Brunner, R., Mahabal, A., Lopes, P., Kohl Moreira, J., & Stalder, B. (2004). The Digitized Second Palomar

Observatory Sky Survey (DPOSS). III. Star-Galaxy Separation, Astron. J. **128**, pp. 3092-3107.

[11] Donalek, C., Mahabal, A., Djorgovski, S. G., Marney, S., Drake, A., Graham, M., Glikman, E., and Williams, R. (2008). New Approaches to Object Classification in Synoptic Sky Surveys, AIP Conf. Ser., **1082**, pp. 252-256.

[12] Djorgovski, S. G., Mahabal, A., Brunner, R., Gal, R., Castro, S., de Carvalho, R., and Odewahn, S. (2001a). Searches for Rare and New Types of Objects, in: Virtual Observatories of the Future, eds. R. Brunner, S.G. Djorgovski, and A. Szalay, A.S.P. Conf. Ser. **225**, pp. 52-63.

[13] Djorgovski, S. G., Brunner, R., Mahabal, A., Odewahn, S., de Carvalho, R., Gal, R., Stolorz, P., Granat, R., Curkendall, D., Jacob, J., and Castro, S. (2001b). Exploration of Large Digital Sky Surveys, in: Mining the Sky, eds. A. J. Banday et al., ESO Astrophysics Symposia, (Berlin: Springer), pp. 305-322.

[14] Djorgovski, S.G., Mahabal, A., Brunner, R., Williams, R., Granat, R., Curkendall, D., Jacob, J., and Stolorz, P. (2001c). Exploration of Parameter Spaces in a Virtual Observatory, in: Astronomical Data Analysis, eds. J.-L. Starck & F. Murtagh, Proc. SPIE **4477**, pp. 43-52.

[15] Mahabal, A., Djorgovski, S. G., Drake, A., Donalek, C., Graham, M., Moghaddam, B., Turmon, M., Williams, R., Beshore, E., and Larson, S. (2011). Discovery, Classification, and Scientific Exploration of Transient Events from the Catalina Real-Time Transient Survey, Bull. Astr. Soc. India, **39**, 387-408.

[16] Djorgovski, S. G., Mahabal, A., Drake, A., Graham, M., Donalek, C., & Williams, R., (2012a). Exploring the Time Domain with Synoptic Sky Surveys, in: Proc. IAU Symp. 285: New Horizons in Time Domain Astronomy, eds. E. Griffin et al., (Cambridge: Cambridge Univ. Press), pp. 141-146.

[17] Djorgovski, S. G., Mahabal, A., Donalek, C., Graham, M., Drake, A., Moghaddam, B., and Turmon, M. (2012b). Flashes in a Star Stream: Automated Classification of Astronomical Transient Events, Proc. IEEE e-Science 2012, (IEEE press), 6404437.

[18] Graham, M., Djorgovski, S. G., Mahabal, A., Donalek, C., Drake, A., and Longo, G. (2012). Data Challenges of Time Domain Astronomy, Distributed and Parallel Databases, **30**, pp. 371-384.

[19] Graham, M., Drake, A., Djorgovski, S. G., Mahabal, A., and Donalek, C. (2017). Challenges in the automated classification of variable stars in large databases, in: Wide-Field Variability Surveys: A 21st Century Perspective, eds. Catelan, M. and Gieren, W., EPJ Web of Conferences, **152**, 03001.

[20] Andreoni, I., Coughlin, M., Kool, E., Kasliwal, M., Kumar, H., Bhalerao, V., Carracedo, A., Ho, A., Pang, P., Saraogi, D., et al. Astrophys. J. **918**, 63 (2021).

[21] Cabero, M., Mahabal, A., and McIver, J., Astrophys. J. Lett. **904**, L9 (2020). doi:10.3847/2041-8213/abc5b5 [arXiv:2010.11829 [gr-qc]].

[22] Abbott, T.C., Buffaz, E., Vieira, N., Cabero, M., Haggard, D., Mahabal, A. and McIver, J., Accepted to Astrophys. J., 2022. GWSkyNet-Multi: A Machine Learning Multi-Class Classifier for LIGO-Virgo Public Alerts. https://arxiv.org/abs/2111.04015.

[23] Donalek, C., Kumar, A., Djorgovski, S.G., Mahabal, A., Graham, M., Fuchs, T., Turmon, M., Sajeeth Philip, N., Yang, T.-C., and Longo, G. (2013). Feature Selection Strategies for Classifying High Dimensional Astronomical Data Sets, in: Scalable Machine Learning: Theory and Applications, IEEE BigData 2013, 35-40.

[24] D'Isanto, A., Cavuoti, S., Brescia, M., Donalek, C., Longo, G., Riccio, G., Djorgovski, S. G. (2016). An Analysis of Feature Relevance in the Classification of Astronomical Transients with Machine Learning Methods, MNRAS **457**, pp. 3119-3132.

[25] van Roestel J., Duev D. A., Mahabal A. A., Coughlin M. W., Mr´oz P., Burdge K., Drake A., et al. (2021). Astron. J, **161**, 267.

[26] Gebru T., Morgenstern J., Vecchione B., Wortman Vaughan J., Wallach H., Daumé H., Crawford, K. (2018). Online at https://arxiv.org/abs/1803.

[27] Mitchell M., Wu S., Zaldivar A., Barnes P., Vasserman, L., Hutchinson B., Spitzer E., et al., 2019, FAT* '19: Conference on Fairness, Accountability, and Transparency, https://arxiv.org/abs/1810.03993.

[28] Mahabal, A., Hare, T., Fox, V., Hallinan, and G. (2021). In-space Data Fusion for More Productive Missions. Bull. AAS, **53** (4).

[29] Caldeira, J. and Nord, B., 2020. Deeply uncertain: comparing methods of uncertainty quantification in deep learning algorithms. Machine Learning: Science and Technology, **2(1)**, 015002.

[30] Martínez-Galarza J. R., Bianco F. B., Crake D., Tirumala K., Mahabal A. A., Graham M. J., Giles D., 2021, MNRAS, **508**, 5734.

[31] Rasmussen, C., and Williams, C., Gaussian Processes for Machine Learning (2006). MIT Press.

[32] Lee, J., Bahri, Y., Novak, R., Schoenholz, S., Pennington, J., Sohl-Dickstein, (2017) Deep neural networks as gaussian processes https://arxiv.org/abs/1711.00165.

[33] Foreman-Mackey, D., Agol, E., Ambikasaran, S., and Angus, R. (2017). Fast and scalable Gaussian process modeling with applications to astronomical time series, Astron. J., **154**, 220.

[34] Tachibana, Y., Graham, M., Kawai, N., Djorgovski, S. G., Drake, A., and Mahabal, A. (2020). Deep Modeling of Quasar Variability, Astrophys. J., **903**, 17.

[35] Abdar, M., Pourpanah, F., Hussain, S., Rezazadegan, D., Liu, L., Ghavamzadeh, M., Fieguth, P., Cao, X., Khosravi, A., Acharya, U.R. and Makarenkov, V. A review of uncertainty quantification in deep learning: Techniques, applications and challenges. *Information Fusion*, **76**, 243-297 (2021).

[36] Walmsley M., Smith L., Lintott C., Gal Y., Bamford S., Dickinson H., Fortson L., et al. (2020). Galaxy Zoo: probabilistic morphology through Bayesian CNNs and active learning, MNRAS, **491**, 1554-1574.

[37] Webb S., Lochner M., Muthukrishna, D., Cooke J., Flynn, C., Mahabal A., Goode S., et al., 2020, MNRAS, **498**, 3077.

[38] Villar, V. A., Cranmer, M., Berger, E., Contardo, G., Ho, S., Hosseinzadeh, G., Lin, J. (2021). A Deep-learning Approach for Live Anomaly Detection of Extragalactic Transients, Astrophys. J. Suppl. Ser., **255**, 24.

[39] Ishida E. E. O., Kornilov M. V., Malanchev K. L., Pruzhinskaya M. V., Volnova A. A., Korolev V. S., Mondon F., et al., 2021, Astron, Astrophys., **650**, A195.

[40] Lochner, M., and Bassett, B. (2021). ASTRONOMALY: Personalised active anomaly detection in astronomical data, Astron. Computing, **36**, 100481.

[41] Ball, N., Brunner, R., Myers, A., Strand, N., Alberts, S., and Tcheng, D. (2008). Robust Machine Learning Applied to Astronomical Data Sets. III. Probabilistic Photometric Redshifts for Galaxies and Quasars, Astrophys. J., **683**, 12-21.

[42] Laurino, O., D'Abrusco, R., Longo, G., and Riccio, G. (2011). Astroinformatics of galaxies and quasars: a new general method for photometric redshifts estimation, MNRAS **418**, 2165-2195.

[43] Brescia, M., Cavuoti, S., D'Abrusco, R., Longo, G., and Mercurio, A. (2013). Photometric Redshifts for Quasars in Multi-band Surveys, Astrophys. J., **772**, 140.

[44] Carrasco Kind, M., and Brunner, R. (2014). Exhausting the information: novel Bayesian combination of photometric redshift PDFs, MNRAS, **442**, 3380-3399.

[45] Cavuoti, S., Brescia, M., Longo, G., and Mercurio, A. (2012). Photometric redshifts with the quasi-Newton algorithm (MLPQNA): Results in the PHAT1 contest, Astron. Astrophys., **546**, A13.

[46] Cavuoti, S., Amaro, V., Brescia, M., Vellucci, C., Tortora, C., and Longo, G. (2017). METAPHOR: A machine-learning-based method for the probability density estimation of photometric redshifts, MNRAS, **465**, 1959-1973.

[47] D'Isanto, A., and Polsterer, K. (2018). Photometric redshift estimation via deep learning. Generalized and pre-classification-less, image based, fully probabilistic redshifts, Astron. Astrophys., **609**, A111

[48] Salvato, M., Ilbert, O., and Hoyle, B. (2019). The many flavours of photometric redshifts, Nature Astronomy, **3**, 212-222.

[49] Schmidt, S.J., et al., LSST Dark Energy Science Collaboration (2020). Evaluation of probabilistic photometric redshift estimation approaches for The Rubin Observatory Legacy Survey of Space and Time (LSST), MNRAS **499**, 1587-1606.

[50] Razim, O., Cavuoti, S., Brescia, M., Riccio, G., Salvato, M., and Longo, G. (2021). Improving the reliability of photometric redshift with machine learning, MNRAS, **507**, 5034-5052.

[51] Krone-Martins, A., Delchambre, L., Wertz, O., Ducourant, C., Mignard, F., Teixeira, R., et al. (2018). Gaia GraL: Gaia DR2 gravitational lens systems. I. New quadruply imaged quasar candidates around known quasars, Astron. Astrophys**., 616**, L11.

[52] Krone-Martins, A., Graham, M., Stern, D., Djorgovski, S. G., Delchambre, L., Ducourant, C., et al. (2022). Gaia GraL: Gaia DR2 Gravitational Lens Systems.V.Doubly-imaged QSOs Discovered from Entropy and Wavelets, Astron. Astrophys. in press.

[53] Kamdar, H., Turk, M., and Brunner, R. (2016). Machine learning and cosmological simulations - II. Hydrodynamical simulations, MNRAS, **457**, 1162-1179.

[54] Ni, Y., Li, Y., Lachance, P., Croft, R., Di Matteo, T., Bird, S., and Feng, Y. (2021). AI-assisted superresolution cosmological simulations - II. Halo substructures, velocities, and higher order statistics, MNRAS, **507**, 1021-1033.

[55] Lovell, C., Wilkins, S., Thomas, P., Schaller, M., Baugh, C., Fabbian, G., Bahé, Y. (2022). A machine learning approach to mapping baryons on to dark matter haloes using the EAGLE and C-EAGLE simulations, MNRAS, **509**, 5046-5061.

[56] Chacón, J., Vázquez, J., and Almaraz, E. (2022). Classification algorithms applied to structure formation simulations, Astron. Computing, **3**8, 100527.

[57] Karniadakis, G., Kevrekidis, I., Lu, L., Perdikaris, P., Wang, S., and Yang, L. (2021). Physics-informed machine learning, Nature Reviews Physics, **3**(6):422–440.

[58] Liu, Z., Chen, Y., Du, Y., and Tegmark, M. (2021). Physics-Augmented Learning: A New Paradigm Beyond Physics-Informed Learning, https://arxiv.org/abs/2109.13901

[59] Graham, M., Djorgovski, S. G., Mahabal, A., Donalek, C., and Drake, A. (2013). Machine-Assisted Discovery of Relationships in Astronomy, MNRAS, **431**, 2371.

[60] Udrescu, S.-M., and Tegmark, M. (2020). AI Feynman: a Physics-Inspired Method for Symbolic Regression, Science Advances, **6**:2631.

[61] Sun, H., and Moscato, P. (2019). A Memetic Algorithm for Symbolic Regression, in: 2019 IEEE Congress on Evolutionary Computation (CEC), 2167-2174.

[62] Moscato, P., Sun, H., and Haque, M. (2021). Analytic Continued Fractions for Regression: A Memetic Algorithm Approach, Expert Systems with Applications, **179**, 115018.

[63] Djorgovski, S. G., Graham, M., Donalek, C., Mahabal, A., Drake, A., Turmon, M., and Fuchs, T. (2016). Real-Time Data Mining of Massive Data Streams from Synoptic Sky Surveys, Future Gen. Comp. Sys., **59**, 95-104.

[64] Djorgovski, S. G., Mahabal, A., Donalek, C., Graham, M., Drake, A., Turmon, M., Fuchs, T. (2014). Automated Real-Time Classification and Decision Making in Massive Data Streams from Synoptic Sky Surveys, Proc. IEEE e-Science 2014, ed. C. Medeiros, (IEEE press), pp. 204-211.

[65] Mahabal, A., Sheth, K., Gieseke, F., Pai, A., Djorgovski, S.G., Drake, A., Graham, M., and CSS/CRTS/PTF Teams (2017). Deep-Learnt Classification of Light Curves, in 2017 IEEE Symp. on Computational Intelligence (SSCI), 2757-2764.

[66] Mahabal, A., et al. (ZTF Team) (2019). Machine Learning for the Zwicky Transient Facility, Publ. Astron. Soc. Pacific **131**, 038002.